\documentclass[aps,prl,twocolumn,floatfix, showpacs]{revtex4-1}
\usepackage{graphicx}
\usepackage{dcolumn}
\usepackage{bm}

\usepackage{array}
\usepackage{subfigure}
\usepackage{epstopdf}

\newcommand{\be}{\begin{equation}}
\newcommand{\ee}{\end{equation}}

\newcommand\Pra{\mbox{\textrm{Pr}}} 
\newcommand\Ra{\mbox{\textrm{Ra}}} 

\def\gtwid{\mathrel{\raise.3ex\hbox{$>$\kern-.75em\lower1ex\hbox{$\sim$}}}}
\def\ltwid{\mathrel{\raise.3ex\hbox{$<$\kern-.75em\lower1ex\hbox{$\sim$}}}}

\begin{document}

\title{Logarithmic temperature profiles in turbulent Rayleigh-B\'enard convection}

\author{Guenter Ahlers$^1$}
\author{Eberhard Bodenschatz$^{2,3,4}$}
\author{Denis Funfschilling$^5$}
\author{Siegfried Grossmann$^6$}
\author{Xiaozhou He$^2$}
\author{Detlef Lohse$^7$}
\author{Richard J.A.M. Stevens$^7$}
\author{Roberto Verzicco$^{7,8}$}

\affiliation{$^1$Department of Physics, University of California, Santa Barbara, CA 93106, USA}
\affiliation{$^2$Max Planck Institute for Dynamics and Self-Organization, D-37073 Goettingen, Germany}
\affiliation{$^{3}$Institute for Nonlinear Dynamics, University of G\"ottingen, D-37073 G\"ottingen, Germany}
\affiliation{$^{4}$Laboratory of Atomic and Solid-State Physics and Sibley School of Mechanical and Aerospace Engineering, Cornell University, Ithaca, New York 14853}
\affiliation{$^5$LSGC CNRS - GROUPE ENSIC, BP 451, 54001 Nancy Cedex, France}
\affiliation{$^6$Fachbereich Physik der Philipps-Universit\"at, Renthof 6, D-35032
Marburg, Germany}
\affiliation{$^7$Department of Science and Technology and J.M. Burgers Center for Fluid Dynamics, University of Twente, P.O Box 217, 7500 AE Enschede, The Netherlands}
\affiliation{$^8$Dept. of Mech. Eng., UniversitaÕ di Roma ÓTor VergataÓ, Via del Politecnico 1, 00133, Roma.}

\date{\today}

\begin{abstract}

We report results for the temperature profiles of turbulent Rayleigh-B\'enard convection (RBC) in the interior of a cylindrical sample of aspect ratio $\Gamma \equiv D/L = 0.50$ ($D$ and $L$ are the diameter and height respectively). Results from experiment over the Rayleigh number range $4\times 10^{12} \alt Ra \alt 10^{15}$ for a Prandtl number $\Pra \simeq 0.8$ and from direct numerical simulation (DNS) at $Ra = 2 \times 10^{12}$ for $\Pra = 0.7$ are presented. We find that the temperature varies as $A*ln(z/L) + B$  where $z$ is the distance from the bottom or top plate. This is the case in the classical as well as in the ultimate state of RBC. From DNS we find that $A$ in the classical state decreases in the radial direction as the distance from the side wall increases and becomes small near the sample center. 

\end{abstract}


\maketitle

Turbulent convection of a fluid contained between two horizontal plates separated by a distance $L$  and heated from below (Rayleigh-B\'enard convection or RBC) \cite{Ah09,AGL09,LX10} is a system in fluid mechanics with many features that are of fundamental interest. It is also a phenomenon with numerous astrophysical \cite{CEW03,Bu94,Nor03}, geophysical \cite{CO94,GCHR99,DDSC00,HMF01,MS99,Ra00}, and technological \cite{Sta10} applications. Nonetheless some of its properties remain poorly explored and understood. A ``classical" state of RBC  exists below a transition range to an ``ultimate" state; the transition range extends over more than a decade from $\Ra^*_{1}$ to $\Ra^*_{2}$ \cite{HFNBA12}  ($\Ra$ is a dimensionless measure of the applied temperature difference). For simplicity  we shall characterize this range by $\Ra^*$ which, for the parameters of our work, is about  $10^{14}$ \cite{GL02,HFNBA12}.
For the classical state it is known from experiment (see, for instance, \cite{TBL93,BTL93,BTL94,XX97,LX98,ZX01,WX04,PRTBT05}) that approximately half of the applied temperature difference $\Delta T$ is sustained by two thin thermal boundary layers (BLs), one just below the top and the other just above the bottom plate. These BLs are of Prandtl-Blasius type, i.e. laminar, albeit fluctuating  \cite{ZX10,SZGVXL12}. The entire interior of the sample, known as the ``bulk", is then approximately isothermal in the time average, but it also undergoes vigorous local temperature fluctuations. At a more detailed level it has long been recognized that the bulk actually sustains small temperature gradients, but these gradients were believed to be more or less  independent of vertical position and the total temperature drop across the bulk  is known to be much smaller than that across the BLs (see, for instance, \cite{TBL93,BA07_EPL,WA11a}).  

We find that, beyond a thin boundary layer unresolved in the experiment, the temperature $T(z)$ and its root-mean-square (rms) fluctuations $\sigma(z)$ vary {\it logarithmically} as a function of the distance $z$ from the bottom plate. While
 a logarithmic dependence was predicted for the ultimate state above $\Ra^*$ where the BLs are turbulent and believed to extend throughout the entire sample \cite{GL11}, to our knowledge there is no theory at present that predicts the bulk properties for $\Ra < \Ra^*$. We believe that the discovery of logarithmic profiles is an important step  toward developing a more fundamental understanding of the bulk. On the one hand, 
its origin may be sought in the diffusion of the excess or deficit of enthalpy carried by plumes.  On the other hand, 
the logarithmic dependence suggests a relationship to the well known logarithmic variation of the velocity in turbulent shear flows discussed originally by von K\'arm\'an \cite{Ka30} and Prandtl \cite{Pr32} (for a recent review, see \cite{MMMNSS10}).

We present experimental measurements for a Prandtl number $\Pra \simeq 0.8$ over the range $4\times 10^{12} \alt \Ra \alt 10^{15}$ and analyze direct numerical simulations (DNS) for $\Pra = 0.7$ and $\Ra = 2\times 10^{12}$ \cite{SLV11} for a cylindrical sample of aspect ratio $\Gamma \equiv D/L =0.50$ ($D$ is the diameter). Both experiment and DNS show that, through most of the bulk, the dimensionless time-averaged temperature $\Theta(z) \equiv [\langle T(z )\rangle - T_m] / (T_b - T_t)$  (we denote the time average by $\langle ... \rangle$ and $T_m \equiv (T_b + T_t)/2$ with $T_b$ and $T_t$ the temperatures at the top and bottom of the sample) can be represented well by 
\be
\Theta(z)  = A*ln(z/L) + B\ .
\label{eq:log}
\ee
We also find that the root-mean-square temperature fluctuations 
$\sigma(z) \equiv \langle [ T(z) - \langle T(z) \rangle ] ^2 \rangle ^{1/2} / (T_b - T_t)$
are consistent with a logarithmic dependence on $z$, and represent them by
\be
\sigma(z) = C*ln(z/L) + D\ .
\label{eq:log2}
\ee
The DNS data show that the amplitude $A(r)$ is largest near the side wall and decreases gradually as the distance $R-r$  from the wall increases ($R = D/2$ is the sample radius and $r$ the radial coordinate). At the sample center $A$ nearly vanishes. The experimental data cover a wide range of \Ra\ but are for a single $(R-r)/L = 0.0045$. This location is, however, well inside the bulk; the DNS data at $\Ra = 2\times 10^{12}$ showed that the viscous BL only extends to $(R-r)/L \approx 0.0008$.

The measurements were made with a large cylindrical sample of height $L = 2.24$ m and diameter $D = 1.12$ m known as the High-Pressure Convection Facility II (HPCF-II) which was placed in an even larger pressure vessel known as the ``Uboot of G\"ottingen" at the Max Planck Institute for Dynamics and Self Organization in G\"ottingen, Germany \cite{AFB09}. The Uboot and HPCF-II were filled with the gas sulfur hexafluoride (SF$_6$). During the measurements the HPCF-II was completely sealed.  The Prandtl number $\Pra \equiv \nu/\kappa$ ($\nu$ is the kinematic viscosity and $\kappa$ the thermal diffusivity)  was 0.79 (0.86) near $\Ra = 4\times 10^{12}$ ($10^{15}$). The measurements were made at $T_m \simeq 21^\circ$C and at various pressures up to 19 bars. The Rayleigh number is given by  $\Ra = \alpha g \Delta T L^3/\kappa \nu$. Here  the isobaric thermal expansion coefficient $\alpha$, as well as $\kappa$ and $\nu$, were evaluated at $T_m$, and $g$ is the acceleration of gravity. Typically $\Delta T$ had values in the range from 4 to 16 K. 

The sample was tilted slightly, with its axis at an angle of 14 mrad relative to gravity. 
We ensured that the tilting had no effect on our results. This is understandable because at these high $Ra$ values no pronounced large scale circulation (LSC) exists \cite{HFNBA12}. 
Two sets of thermistors were installed for the temperature-profile measurements. One was located at what would be the preferred down-flow orientation at lower $Ra$, and the other was removed from the first  in the azimuthal direction by an angle $\pi$. Each set consisted of eight thermometers which were located in the fluid $1.0 \pm 0.1$ cm from the side wall, {\it i.e.} at a radial position $(R-r)/L = 0.00445 \pm 0.0005$. The eight thermistors were located at  $z = 4.0, 6.1, 8.1, 12.1, 16.1, 32.2, 64.2$, and 110.5 cm, with an uncertainty of the vertical position of 0.1 cm. 

\begin{figure}
\includegraphics[width=3.25in]{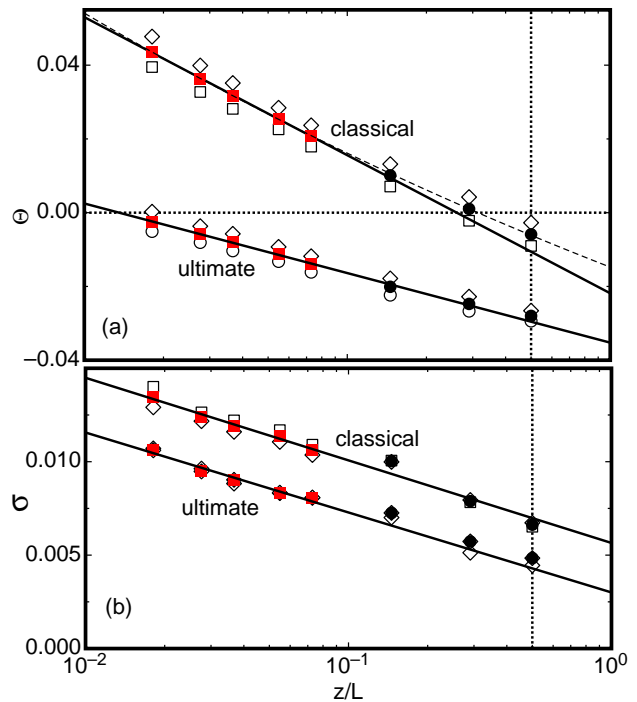}
\caption{The measurements are for a radial position $(R-r)/L = 0.0045$. (a): The time-averaged temperature $\Theta(z) \equiv [\langle T(z)\rangle - T_m] / (T_b - T_t)$ as a function of the vertical position $z/L$. The vertical dotted line marks the sample center at $z/L = 0.5$. The open squares (diamonds) are results at the preferred down-flow (up-flow) orientation of the LSC. The solid symbols are their averages. The solid squares (red online, $z/L \alt 0.08$) were used for the fits of Eq.~\ref{eq:log} to the data. The solid lines are those fits. The upper data set is in the classical regime for $\Ra = 1.18\times 10^{13}$ and $\Pra = 0.787$. It gave $A = -0.0162 \pm 0.0004$ and $B = -0.00218 \pm 0.0007$ where the uncertainties are the standard errors obtained from the fit. The dashed line is a fit of a power law $\Theta(z) = \tilde A*(z/L)^\zeta + \tilde B$ to the data with $z/L \alt 0.08$ which yielded $\zeta = -0.09 \pm 0.05$. The lower data set is for the ultimate state at $\Ra = 9.64\times 10^{14}$ and $\Pra = 0.862$. It gave $A = -0.0082 \pm 0.00016$ and $B = -0.0352 \pm 0.0005$.
(b): The root-mean-square fluctuations $\sigma(z) = \langle [ T(z) - \langle T(z) \rangle ] ^2 \rangle ^{1/2} / (T_b - T_t)$ as a function of $z/L$ corresponding to the data in (a). The symbols are as in (a), and the fit parameters  of Eq.~\ref{eq:log2} are $C = -0.00192\pm 0.00013$ and $D = 0.00565\pm 0.00044$ for the upper (classical) set at $\Ra = 1.18\times 10^{13}$ and  $C = 0.0019\pm 0.0002$ and $D = 0.0030\pm 0.0007$ for the lower (ultimate) set at $\Ra = 9.64\times 10^{14}$.}
\label{fig:profile1}
\end{figure}

Two typical measurements of the sixteen time-averaged temperatures are shown in Fig.~\ref{fig:profile1}a as a function of $z/L$ on a logarithmic scale. The top (bottom) one is for $\Ra =  1.18\times 10^{13}$ ($9.64\times 10^{14}$) in the classical (ultimate) state. For each case one sees that the data sets at the two azimuthal orientations differ slightly from each other. We attribute this to the influence of remnants of the LSC on the temperature profiles \cite{BNA05}. To compensate for this effect, we henceforth consider the average at each vertical position of the two data sets, as shown by the solid symbols in the figure. Except at the largest $z/L$, these data fall on straight lines and thus are represented well by Eq.~\ref{eq:log} for more than a decade.
 
In a sample that conforms perfectly to the Boussinesq approximation we would expect another logarithmic dependence emanating from the top plate to meet the data shown in the figure at $T_m$ ({\it i.e.} at $\Theta = 0$) and $z/L = 1/2$. However, in the experiment we find that $\Theta(z/L=1/2) < 0$, albeit only by 0.006 (0.028) for $Ra=1.18\times10^{13}$ ($9.64 \times 10^{14}$). We do not know the reason for this offset. However, it will necessarily lead to a small departure from the logarithmic dependence because the two branches, one coming from the bottom and the other from the top plate, must form an analytic function with a continuous derivative at $z/L = 1/2$ where they meet.
For a quantitative analysis we therefore fit Eq.~\ref{eq:log} only to the five points with $z/L \alt 0.08$. The resulting functions are shown as the solid lines in the figure. 

It should be mentioned that, as for any log-behavior in the limited range of 1.5 decades, 
one can also obtain a good fit to the data 
with the power law $\Theta(z) = \tilde A*(z/L)^\zeta + \tilde B$. This is shown by the dashed line in Fig.~\ref{fig:profile1}a. However, the resulting exponent $\zeta = 0.09 \pm 0.05$ is quite small and unlike any other known exponent relevant to this system \cite{AGL09,LX10}. Further, its uncertainty  is not much smaller than its value. Thus we shall continue the presentation of the results in terms of the logarithmic function Eq.~\ref{eq:log}.

The rms temperature fluctuations $\sigma$ are shown in Fig.~\ref{fig:profile1}b. They too are described well by the logarithmic form. Also in this case the relevant equation (Eq.~\ref{eq:log2}) was fit to the data only for $z/L \alt 0.08$ to determine $C$ and $D$.

\begin{figure}
\includegraphics[width=3in]{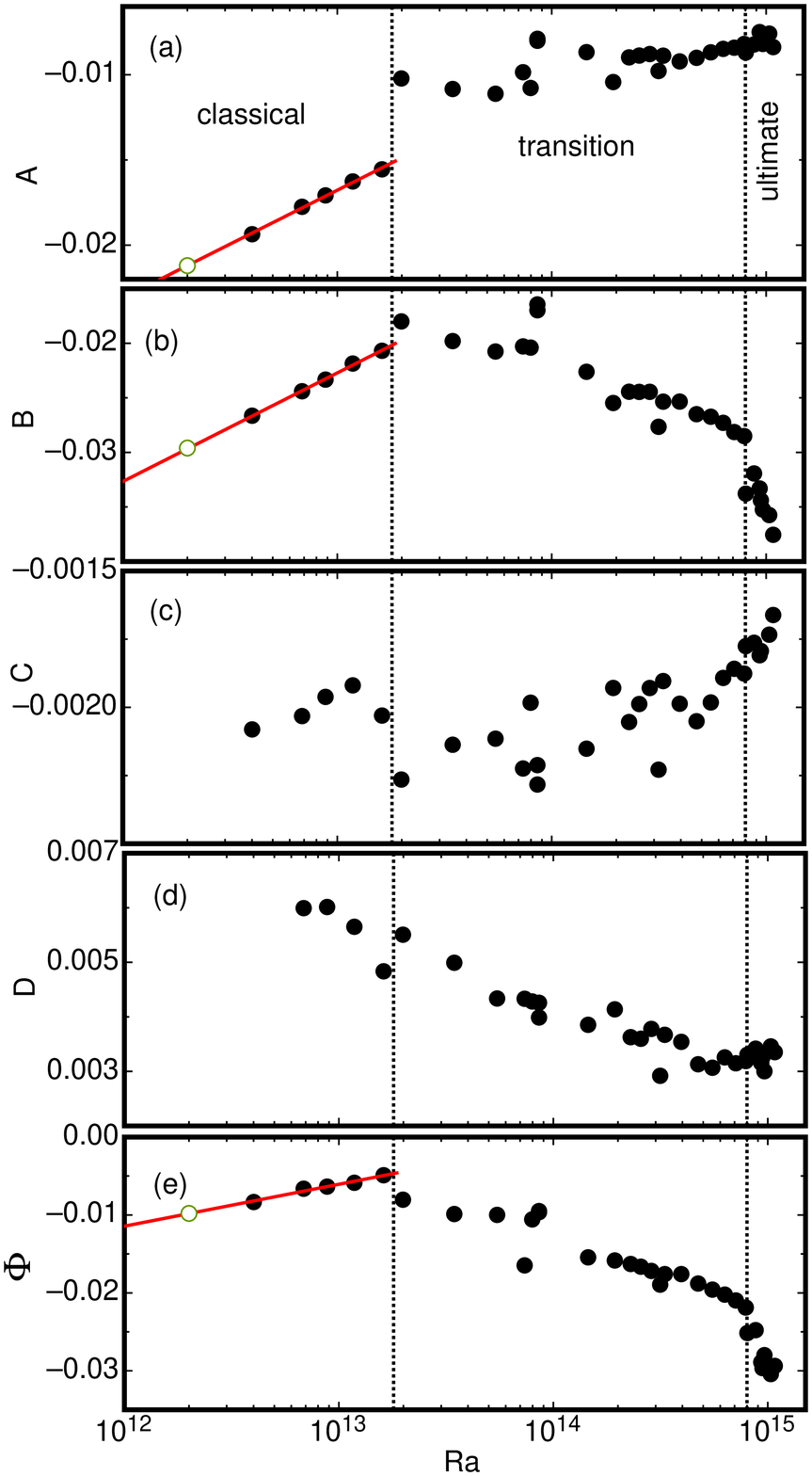}
\caption{The parameters $A, B, C$, and $D$ obtained by fitting Eqs.~\ref{eq:log} and \ref{eq:log2} to the experimental temperature and fluctuation profiles, and the deviation $\phi \equiv (T_c - T_m)/\Delta T$ of the temperature $T_c = T(z/L = 1/2)$ from the mean temperature $T_m$, all as a function of \Ra. All data are for a radial position $(R - r)/L = 0.0045$. The vertical dotted lines indicate the locations of $\Ra_1^*$ and $\Ra_2^*$. The solid lines (red online) are fits of the function $A = a*log_{10}(\Ra) + b$  to the data with $\Ra < \Ra_1^*$. The extrapolation to $\Ra = 2\times 10^{12}$ (open circles, green online) yielded $A = -0.0212,~B = -0.0296,$ and $\Phi = -0.0098$.}
\label{fig:parameters}
\end{figure}

The parameters $A, B, C,$ and $D$, as well as the deviation of the center temperature $T_c = T(z/L=0.50)$ from $T_m$,  are shown in Fig.~\ref{fig:parameters} as a function of \Ra. As had been reported before for measurements of the Nusselt and the Reynolds number \cite{HFNBA12} and reported above, there is a  {\it range} of \Ra\ which extends from $\Ra_1^*$ to $\Ra_2^*$ (the vertical dotted lines in the figure) over which the transition from the classical to the ultimate state takes place. The locations of $\Ra_1^*$ and $\Ra_2^*$ are particularly noticeable in the data for $B$ and $\Phi$. In the transition region the parameters scatter much more than above or below it because the state assumed by the system can vary from one experimental point to another.

\begin{figure}
 \includegraphics[width=3in]{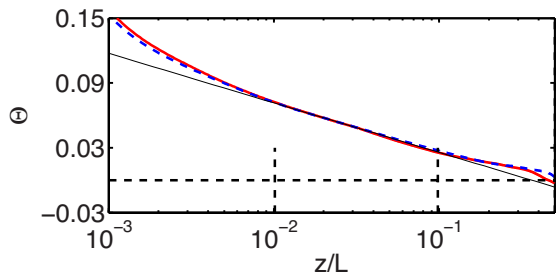}
\caption{ The temperature $\Theta(z)$ as a function of $z/L$ at a radial position $(R-r)/L = 0.0045$ from DNS for $\Ra = 2\times 10^{12}$, $\Pra = 0.7$, and $\Gamma = 1/2$. The solid red (dashed blue) line indicates the profile measured from the bottom (top) plate as a function of $z/L$ (here $z$ is taken as the distance from the nearest plate). The thin line represents a fit of Eq.~\ref{eq:log} over the range $10^{-2}\leq z/L \leq 0.1$ to the average of the two profiles (the fitting interval is indicated by the two short dashed vertical lines).}
\label{fig:DNS1}
\end{figure}

\begin{figure}
\includegraphics[width=3.5in]{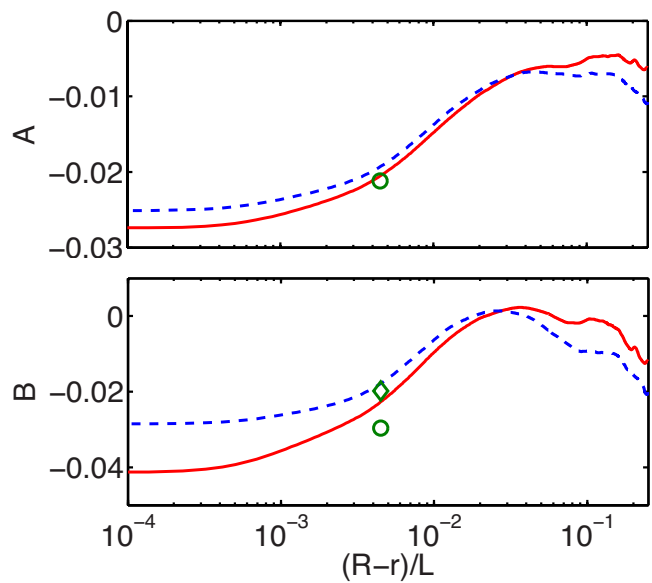}
\caption{DNS results for $A$ and $B$ in Eq.~\ref{eq:log} as a function of $(R - r)/L$ on a logarithmic scale. The solid red (dashed blue) lines indicate the result obtained above  the bottom (below the top) plate. The open circles (green online) correspond to the extrapolations of the experimental data for $A$ and $B$ as shown in Figs.~\ref{fig:parameters}a and b to $\Ra = 2\times 10^{12}$. The open diamond indicates the value of $B$ corrected for $\Phi$, see Fig.~\ref{fig:parameters}e.}
\label{fig:DNS2}
\end{figure}

In Fig.\ \ref{fig:DNS1} we show the results for $\Theta(z)$ obtained from DNS at $\Ra = 2\times 10^{12}$ and $\Pra = 0.7$ \cite{SLV11}. Those profiles are for  $(R-r)/L = 0.0045$, i.e. the same radial position as that of the experiment, and are based on azimuthally and time averaged temperature data. In this (and the following) figure we show the measured profiles in the simulations both in the top and the bottom half of the sample. The difference between these two is an indication of the statistical uncertainty due to limited time averaging, 
which of course is a more serious limitation in numerics than it is in the experiment.  
Just as for the experimental data, the profile at larger $z/L$ can be described well by Eq.~(\ref{eq:log}).  Figure~\ref{fig:DNS2} gives the DNS results for $A$ and $B$ as a function of the radial position $(R - r)/L$, based on the temperature data for $10^{-2} \leq z/L \leq 10^{-1}$. In this figure one sees that there is excellent agreement between the values of $A$ and $B$ (when corrected for the temperature measured at $z/L=0.50$ (Fig.~\ref{fig:parameters}e)) measured in the experiment and simulations. In addition, the figure reveals that the magnitude of $A$ is largest near the side wall and that it decreases (approximately logarithmically) as the sample interior is approached. Although it remains finite even at the sample centerline [$(R - r)/L = 1/4$], one might conjecture that it would indeed vanish in a sample of larger aspect ratio. Thus the DNS indicates that the logarithmic vertical temperature profile is related to the existence of the side wall; however, it penetrates radially deep into the sample, over a distance that is well over an order of magnitude larger than the BL thickness.

In this Letter we reported on results obtained by using a combination of experiment and DNS to study the interior of turbulent RBC. For the classical state, which exists for $\Ra < \Ra^* \simeq 10^{14}$ and which has laminar BLs adjacent to the top and bottom plates, we find that the bulk which is found between these two layers sustains a non-trivial and interesting temperature field $\Theta(z/L,r/L)$. Whereas it had generally been assumed that the temperature in the sample interior is either constant or varying linearly and slowly in space, we find that $\Theta$ varies logarithmically with distance from the plates over a wide range of $z/L$. The root-mean-square temperature fluctuations show similar variations. The amplitude of the logarithmic profile is largest near the side wall and becomes small as  the sample center is approached. The origin of the logarithmic profile remains unclear. On the one hand one may speculate that it is the result of the diffusion of enthalpy carried from the BLs into the interior by plumes; but a model for this process which would yield a logarithmic distribution is not known to us. On the other hand, the logarithmic variation suggests a possible relationship to the well known logarithmic velocity profiles in shear flows \cite{Ka30,Pr32,MMMNSS10}.

In the ultimate state, which exists near $\Ra = 10^{15}$ and above \cite{GL02,HFNBA12}, it was predicted  \cite{GL11} 
that  the BLs are turbulent and that they extend vertically throughout the entire sample; thus there is no ``bulk" in the same sense as there is for the classical state. A logarithmic temperature profile due to the turbulent BLs was predicted to extend from each plate deep into the sample, with the two profiles meeting at half height. Indeed,
 the experimental measurements in the ultimate state do find a logarithmic  dependence of the temperature on the vertical coordinate. Unfortunately,
  these large values of \Ra\ are not yet accessible to DNS (and will not be for some time), and experimental results are available only for one radial position. Thus it is not known whether the logarithmic variation persist throughout the sample, as one would expect on the basis of the prediction.

{\it Acknowledgements:}
We are grateful to the Max-Planck Society and the Volkswagen Stiftung for their support of the experiment. We thank the Deutsche Forschungsgemeinschaft (DFG) for financial support through SFB963: ``Astrophysical Flow Instabilities and Turbulence". The work of  G.A. was supported in part by the U.S. National Science Foundation through Grant DMR11-58514. The simulation at $Ra=2\times10^{12}$ was performed as part of a large scale computing project at HLRS (High Performance Computing Center Stuttgart). RJAMS  thanks the Foundation for Fundamental Research on Matter (FOM) for financial support. etc. 


\begin{thebibliography}{34}%
\makeatletter
\providecommand \@ifxundefined [1]{%
 \@ifx{#1\undefined}
}%
\providecommand \@ifnum [1]{%
 \ifnum #1\expandafter \@firstoftwo
 \else \expandafter \@secondoftwo
 \fi
}%
\providecommand \@ifx [1]{%
 \ifx #1\expandafter \@firstoftwo
 \else \expandafter \@secondoftwo
 \fi
}%
\providecommand \natexlab [1]{#1}%
\providecommand \enquote  [1]{``#1''}%
\providecommand \bibnamefont  [1]{#1}%
\providecommand \bibfnamefont [1]{#1}%
\providecommand \citenamefont [1]{#1}%
\providecommand \href@noop [0]{\@secondoftwo}%
\providecommand \href [0]{\begingroup \@sanitize@url \@href}%
\providecommand \@href[1]{\@@startlink{#1}\@@href}%
\providecommand \@@href[1]{\endgroup#1\@@endlink}%
\providecommand \@sanitize@url [0]{\catcode `\\12\catcode `\$12\catcode
  `\&12\catcode `\#12\catcode `\^12\catcode `\_12\catcode `\%12\relax}%
\providecommand \@@startlink[1]{}%
\providecommand \@@endlink[0]{}%
\providecommand \url  [0]{\begingroup\@sanitize@url \@url }%
\providecommand \@url [1]{\endgroup\@href {#1}{\urlprefix }}%
\providecommand \urlprefix  [0]{URL }%
\providecommand \Eprint [0]{\href }%
\@ifxundefined \urlstyle {%
  \providecommand \doi  [0]{\begingroup \@sanitize@url \@doi}%
  \providecommand \@doi [1]{\endgroup \@@startlink {\doibase
  #1}doi:\discretionary {}{}{}#1\@@endlink }%
}{%
  \providecommand \doi  [0]{doi:\discretionary{}{}{}\begingroup
  \urlstyle{rm}\Url }%
}%
\providecommand \doibase [0]{http://dx.doi.org/}%
\providecommand \Doi [0]{\begingroup \@sanitize@url \@Doi }%
\providecommand \@Doi  [1]{\endgroup\@@startlink{\doibase#1}\@@Doi}%
\providecommand \@@Doi [1]{#1\@@endlink}%
\providecommand \selectlanguage [0]{\@gobble}%
\providecommand \bibinfo  [0]{\@secondoftwo}%
\providecommand \bibfield  [0]{\@secondoftwo}%
\providecommand \translation [1]{[#1]}%
\providecommand \BibitemOpen [0]{}%
\providecommand \bibitemStop [0]{}%
\providecommand \bibitemNoStop [0]{.\EOS\space}%
\providecommand \EOS [0]{\spacefactor3000\relax}%
\providecommand \BibitemShut  [1]{\csname bibitem#1\endcsname}%
\bibitem [{\citenamefont {Ahlers}(2009)}]{Ah09}%
  \BibitemOpen
  \bibfield  {author} {\bibinfo {author} {\bibfnamefont {G.}~\bibnamefont
  {Ahlers}},\ }\href@noop {} {\bibfield  {journal} {\bibinfo  {journal}
  {Physics},\ }\textbf {\bibinfo {volume} {2}},\ \bibinfo {pages} {74}
  (\bibinfo {year} {2009})}\BibitemShut {NoStop}%
\bibitem [{\citenamefont {Ahlers}\ \emph
  {et~al.}(2009){\natexlab{a}}\citenamefont {Ahlers}, \citenamefont
  {Grossmann},\ and\ \citenamefont {Lohse}}]{AGL09}%
  \BibitemOpen
  \bibfield  {author} {\bibinfo {author} {\bibfnamefont {G.}~\bibnamefont
  {Ahlers}}, \bibinfo {author} {\bibfnamefont {S.}~\bibnamefont {Grossmann}}, \
  and\ \bibinfo {author} {\bibfnamefont {D.}~\bibnamefont {Lohse}},\
  }\href@noop {} {\bibfield  {journal} {\bibinfo  {journal} {Rev. Mod. Phys.},\
  }\textbf {\bibinfo {volume} {81}},\ \bibinfo {pages} {503} (\bibinfo {year}
  {2009}{\natexlab{a}})}\BibitemShut {NoStop}%
\bibitem [{\citenamefont {Lohse}\ and\ \citenamefont {Xia}(2010)}]{LX10}%
  \BibitemOpen
  \bibfield  {author} {\bibinfo {author} {\bibfnamefont {D.}~\bibnamefont
  {Lohse}}\ and\ \bibinfo {author} {\bibfnamefont {K.-Q.}\ \bibnamefont
  {Xia}},\ }\href@noop {} {\bibfield  {journal} {\bibinfo  {journal} {Annu.
  Rev. Fluid Mech.},\ }\textbf {\bibinfo {volume} {42}},\ \bibinfo {pages}
  {335} (\bibinfo {year} {2010})}\BibitemShut {NoStop}%
\bibitem [{\citenamefont {Cattaneo}\ \emph {et~al.}(2003)\citenamefont
  {Cattaneo}, \citenamefont {Emonet},\ and\ \citenamefont {Weiss}}]{CEW03}%
  \BibitemOpen
  \bibfield  {author} {\bibinfo {author} {\bibfnamefont {F.}~\bibnamefont
  {Cattaneo}}, \bibinfo {author} {\bibfnamefont {T.}~\bibnamefont {Emonet}}, \
  and\ \bibinfo {author} {\bibfnamefont {N.}~\bibnamefont {Weiss}},\
  }\href@noop {} {\bibfield  {journal} {\bibinfo  {journal} {Astrophys. J.},\
  }\textbf {\bibinfo {volume} {588}},\ \bibinfo {pages} {1183} (\bibinfo {year}
  {2003})}\BibitemShut {NoStop}%
\bibitem [{\citenamefont {Busse}(1994)}]{Bu94}%
  \BibitemOpen
  \bibfield  {author} {\bibinfo {author} {\bibfnamefont {F.~H.}\ \bibnamefont
  {Busse}},\ }\href@noop {} {\bibfield  {journal} {\bibinfo  {journal}
  {Chaos},\ }\textbf {\bibinfo {volume} {4}},\ \bibinfo {pages} {123} (\bibinfo
  {year} {1994})}\BibitemShut {NoStop}%
\bibitem [{\citenamefont {Nordlund}(2003)}]{Nor03}%
  \BibitemOpen
  \bibfield  {author} {\bibinfo {author} {\bibfnamefont {A.}~\bibnamefont
  {Nordlund}},\ }\href@noop {} {\emph {\bibinfo {title} {Solar photosphere and
  convection}}}\ (\bibinfo  {publisher} {Cambridge University press},\ \bibinfo
  {address} {Cambridge},\ \bibinfo {year} {2003})\BibitemShut {NoStop}%
\bibitem [{\citenamefont {Cardin}\ and\ \citenamefont {Olson}(1994)}]{CO94}%
  \BibitemOpen
  \bibfield  {author} {\bibinfo {author} {\bibfnamefont {P.}~\bibnamefont
  {Cardin}}\ and\ \bibinfo {author} {\bibfnamefont {P.}~\bibnamefont {Olson}},\
  }\href@noop {} {\bibfield  {journal} {\bibinfo  {journal} {Phys. of the Earth
  and Planetary Interiors},\ }\textbf {\bibinfo {volume} {82}},\ \bibinfo
  {pages} {235} (\bibinfo {year} {1994})}\BibitemShut {NoStop}%
\bibitem [{\citenamefont {Glatzmaier}\ \emph {et~al.}(1999)\citenamefont
  {Glatzmaier}, \citenamefont {Coe}, \citenamefont {Hongre},\ and\
  \citenamefont {Roberts}}]{GCHR99}%
  \BibitemOpen
  \bibfield  {author} {\bibinfo {author} {\bibfnamefont {G.}~\bibnamefont
  {Glatzmaier}}, \bibinfo {author} {\bibfnamefont {R.}~\bibnamefont {Coe}},
  \bibinfo {author} {\bibfnamefont {L.}~\bibnamefont {Hongre}}, \ and\ \bibinfo
  {author} {\bibfnamefont {P.}~\bibnamefont {Roberts}},\ }\href@noop {}
  {\bibfield  {journal} {\bibinfo  {journal} {Nature(London)},\ }\textbf
  {\bibinfo {volume} {401}},\ \bibinfo {pages} {885} (\bibinfo {year}
  {1999})}\BibitemShut {NoStop}%
\bibitem [{\citenamefont {van Doorn}\ \emph {et~al.}(2000)\citenamefont {van
  Doorn}, \citenamefont {Dhruva}, \citenamefont {Sreenivasan},\ and\
  \citenamefont {Cassella}}]{DDSC00}%
  \BibitemOpen
  \bibfield  {author} {\bibinfo {author} {\bibfnamefont {E.}~\bibnamefont {van
  Doorn}}, \bibinfo {author} {\bibfnamefont {B.}~\bibnamefont {Dhruva}},
  \bibinfo {author} {\bibfnamefont {K.~R.}\ \bibnamefont {Sreenivasan}}, \ and\
  \bibinfo {author} {\bibfnamefont {V.}~\bibnamefont {Cassella}},\ }\href@noop
  {} {\bibfield  {journal} {\bibinfo  {journal} {Phys. Fluids},\ }\textbf
  {\bibinfo {volume} {12}},\ \bibinfo {pages} {1529} (\bibinfo {year}
  {2000})}\BibitemShut {NoStop}%
\bibitem [{\citenamefont {Hartmann}\ \emph {et~al.}(2001)\citenamefont
  {Hartmann}, \citenamefont {Moy},\ and\ \citenamefont {Fu}}]{HMF01}%
  \BibitemOpen
  \bibfield  {author} {\bibinfo {author} {\bibfnamefont {D.~L.}\ \bibnamefont
  {Hartmann}}, \bibinfo {author} {\bibfnamefont {L.~A.}\ \bibnamefont {Moy}}, \
  and\ \bibinfo {author} {\bibfnamefont {Q.}~\bibnamefont {Fu}},\ }\href@noop
  {} {\bibfield  {journal} {\bibinfo  {journal} {J. Climate},\ }\textbf
  {\bibinfo {volume} {14}},\ \bibinfo {pages} {4495} (\bibinfo {year}
  {2001})}\BibitemShut {NoStop}%
\bibitem [{\citenamefont {Marshall}\ and\ \citenamefont {Schott}(1999)}]{MS99}%
  \BibitemOpen
  \bibfield  {author} {\bibinfo {author} {\bibfnamefont {J.}~\bibnamefont
  {Marshall}}\ and\ \bibinfo {author} {\bibfnamefont {F.}~\bibnamefont
  {Schott}},\ }\href@noop {} {\bibfield  {journal} {\bibinfo  {journal} {Rev.
  Geophys.},\ }\textbf {\bibinfo {volume} {37}},\ \bibinfo {pages} {1}
  (\bibinfo {year} {1999})}\BibitemShut {NoStop}%
\bibitem [{\citenamefont {Rahmstorf}(2000)}]{Ra00}%
  \BibitemOpen
  \bibfield  {author} {\bibinfo {author} {\bibfnamefont {S.}~\bibnamefont
  {Rahmstorf}},\ }\href@noop {} {\bibfield  {journal} {\bibinfo  {journal}
  {Climate Change},\ }\textbf {\bibinfo {volume} {46}},\ \bibinfo {pages} {247}
  (\bibinfo {year} {2000})}\BibitemShut {NoStop}%
\bibitem [{\citenamefont {Stacey}(2010)}]{Sta10}%
  \BibitemOpen
  \bibfield  {author} {\bibinfo {author} {\bibfnamefont {W.~M.}\ \bibnamefont
  {Stacey}},\ }\href@noop {} {\emph {\bibinfo {title} {Fusion: An Introduction
  to the Physics and Technology of Magnetic Confinement Fusion}}}\ (\bibinfo
  {publisher} {Wiley},\ \bibinfo {address} {New York},\ \bibinfo {year}
  {2010})\BibitemShut {NoStop}%
\bibitem [{\citenamefont {He}\ \emph {et~al.}(2012)\citenamefont {He},
  \citenamefont {Funfschilling}, \citenamefont {Nobach}, \citenamefont
  {Bodenschatz},\ and\ \citenamefont {Ahlers}}]{HFNBA12}%
  \BibitemOpen
  \bibfield  {author} {\bibinfo {author} {\bibfnamefont {X.}~\bibnamefont
  {He}}, \bibinfo {author} {\bibfnamefont {D.}~\bibnamefont {Funfschilling}},
  \bibinfo {author} {\bibfnamefont {H.}~\bibnamefont {Nobach}}, \bibinfo
  {author} {\bibfnamefont {E.}~\bibnamefont {Bodenschatz}}, \ and\ \bibinfo
  {author} {\bibfnamefont {G.}~\bibnamefont {Ahlers}},\ }\href@noop {}
  {\bibfield  {journal} {\bibinfo  {journal} {Phys. Rev. Lett.},\ }\textbf
  {\bibinfo {volume} {108}},\ \bibinfo {pages} {024502} (\bibinfo {year}
  {2012})}\BibitemShut {NoStop}%
\bibitem [{\citenamefont {Grossmann}\ and\ \citenamefont {Lohse}(2002)}]{GL02}%
  \BibitemOpen
  \bibfield  {author} {\bibinfo {author} {\bibfnamefont {S.}~\bibnamefont
  {Grossmann}}\ and\ \bibinfo {author} {\bibfnamefont {D.}~\bibnamefont
  {Lohse}},\ }\href@noop {} {\bibfield  {journal} {\bibinfo  {journal} {Phys.
  Rev. E},\ }\textbf {\bibinfo {volume} {66}},\ \bibinfo {pages} {016305}
  (\bibinfo {year} {2002})}\BibitemShut {NoStop}%
\bibitem [{\citenamefont {Tilgner}\ \emph {et~al.}(1993)\citenamefont
  {Tilgner}, \citenamefont {Belmonte},\ and\ \citenamefont
  {Libchaber}}]{TBL93}%
  \BibitemOpen
  \bibfield  {author} {\bibinfo {author} {\bibfnamefont {A.}~\bibnamefont
  {Tilgner}}, \bibinfo {author} {\bibfnamefont {A.}~\bibnamefont {Belmonte}}, \
  and\ \bibinfo {author} {\bibfnamefont {A.}~\bibnamefont {Libchaber}},\
  }\href@noop {} {\bibfield  {journal} {\bibinfo  {journal} {Phys. Rev. E},\
  }\textbf {\bibinfo {volume} {47}},\ \bibinfo {pages} {R2253} (\bibinfo {year}
  {1993})}\BibitemShut {NoStop}%
\bibitem [{\citenamefont {Belmonte}\ \emph {et~al.}(1993)\citenamefont
  {Belmonte}, \citenamefont {Tilgner},\ and\ \citenamefont
  {Libchaber}}]{BTL93}%
  \BibitemOpen
  \bibfield  {author} {\bibinfo {author} {\bibfnamefont {A.}~\bibnamefont
  {Belmonte}}, \bibinfo {author} {\bibfnamefont {A.}~\bibnamefont {Tilgner}}, \
  and\ \bibinfo {author} {\bibfnamefont {A.}~\bibnamefont {Libchaber}},\
  }\href@noop {} {\bibfield  {journal} {\bibinfo  {journal} {Phys. Rev.
  Lett.},\ }\textbf {\bibinfo {volume} {70}},\ \bibinfo {pages} {4067}
  (\bibinfo {year} {1993})}\BibitemShut {NoStop}%
\bibitem [{\citenamefont {Belmonte}\ \emph {et~al.}(1994)\citenamefont
  {Belmonte}, \citenamefont {Tilgner},\ and\ \citenamefont
  {Libchaber}}]{BTL94}%
  \BibitemOpen
  \bibfield  {author} {\bibinfo {author} {\bibfnamefont {A.}~\bibnamefont
  {Belmonte}}, \bibinfo {author} {\bibfnamefont {A.}~\bibnamefont {Tilgner}}, \
  and\ \bibinfo {author} {\bibfnamefont {A.}~\bibnamefont {Libchaber}},\
  }\href@noop {} {\bibfield  {journal} {\bibinfo  {journal} {Phys. Rev. E},\
  }\textbf {\bibinfo {volume} {50}},\ \bibinfo {pages} {269} (\bibinfo {year}
  {1994})}\BibitemShut {NoStop}%
\bibitem [{\citenamefont {Xin}\ and\ \citenamefont {Xia}(1997)}]{XX97}%
  \BibitemOpen
  \bibfield  {author} {\bibinfo {author} {\bibfnamefont {Y.~B.}\ \bibnamefont
  {Xin}}\ and\ \bibinfo {author} {\bibfnamefont {K.-Q.}\ \bibnamefont {Xia}},\
  }\href@noop {} {\bibfield  {journal} {\bibinfo  {journal} {Phys. Rev. E},\
  }\textbf {\bibinfo {volume} {56}},\ \bibinfo {pages} {3010} (\bibinfo {year}
  {1997})}\BibitemShut {NoStop}%
\bibitem [{\citenamefont {Lui}\ and\ \citenamefont {Xia}(1998)}]{LX98}%
  \BibitemOpen
  \bibfield  {author} {\bibinfo {author} {\bibfnamefont {S.~L.}\ \bibnamefont
  {Lui}}\ and\ \bibinfo {author} {\bibfnamefont {K.-Q.}\ \bibnamefont {Xia}},\
  }\href@noop {} {\bibfield  {journal} {\bibinfo  {journal} {Phys. Rev. E},\
  }\textbf {\bibinfo {volume} {57}},\ \bibinfo {pages} {5494} (\bibinfo {year}
  {1998})}\BibitemShut {NoStop}%
\bibitem [{\citenamefont {Zhou}\ and\ \citenamefont {Xia}(2001)}]{ZX01}%
  \BibitemOpen
  \bibfield  {author} {\bibinfo {author} {\bibfnamefont {S.~Q.}\ \bibnamefont
  {Zhou}}\ and\ \bibinfo {author} {\bibfnamefont {K.-Q.}\ \bibnamefont {Xia}},\
  }\href@noop {} {\bibfield  {journal} {\bibinfo  {journal} {Phys. Rev.
  Lett.},\ }\textbf {\bibinfo {volume} {87}},\ \bibinfo {pages} {064501}
  (\bibinfo {year} {2001})}\BibitemShut {NoStop}%
\bibitem [{\citenamefont {Wang}\ and\ \citenamefont {Xia}(2004)}]{WX04}%
  \BibitemOpen
  \bibfield  {author} {\bibinfo {author} {\bibfnamefont {J.}~\bibnamefont
  {Wang}}\ and\ \bibinfo {author} {\bibfnamefont {K.-Q.}\ \bibnamefont {Xia}},\
  }\href@noop {} {\bibfield  {journal} {\bibinfo  {journal} {Eur. Phys. J. B},\
  }\textbf {\bibinfo {volume} {32}},\ \bibinfo {pages} {127} (\bibinfo {year}
  {2004})}\BibitemShut {NoStop}%
\bibitem [{\citenamefont {du~Puits}\ \emph {et~al.}(2007)\citenamefont
  {du~Puits}, \citenamefont {Resagk}, \citenamefont {Tilgner}, \citenamefont
  {Busse},\ and\ \citenamefont {Thess}}]{PRTBT05}%
  \BibitemOpen
  \bibfield  {author} {\bibinfo {author} {\bibfnamefont {R.}~\bibnamefont
  {du~Puits}}, \bibinfo {author} {\bibfnamefont {C.}~\bibnamefont {Resagk}},
  \bibinfo {author} {\bibfnamefont {A.}~\bibnamefont {Tilgner}}, \bibinfo
  {author} {\bibfnamefont {F.~H.}\ \bibnamefont {Busse}}, \ and\ \bibinfo
  {author} {\bibfnamefont {A.}~\bibnamefont {Thess}},\ }\href@noop {}
  {\bibfield  {journal} {\bibinfo  {journal} {J. Fluid Mech.},\ }\textbf
  {\bibinfo {volume} {572}},\ \bibinfo {pages} {231} (\bibinfo {year}
  {2007})}\BibitemShut {NoStop}%
\bibitem [{\citenamefont {Zhou}\ and\ \citenamefont {Xia}(2010)}]{ZX10}%
  \BibitemOpen
  \bibfield  {author} {\bibinfo {author} {\bibfnamefont {Q.}~\bibnamefont
  {Zhou}}\ and\ \bibinfo {author} {\bibfnamefont {K.-Q.}\ \bibnamefont {Xia}},\
  }\href@noop {} {\bibfield  {journal} {\bibinfo  {journal} {Phys. Rev.
  Lett.},\ }\textbf {\bibinfo {volume} {104}},\ \bibinfo {pages} {104301}
  (\bibinfo {year} {2010})}\BibitemShut {NoStop}%
\bibitem [{\citenamefont {Stevens}\ \emph {et~al.}(2012)\citenamefont
  {Stevens}, \citenamefont {Zhou}, \citenamefont {Grossmann}, \citenamefont
  {Verzicco}, \citenamefont {Xia},\ and\ \citenamefont {Lohse}}]{SZGVXL12}%
  \BibitemOpen
  \bibfield  {author} {\bibinfo {author} {\bibfnamefont {R.~J. A.~M.}\
  \bibnamefont {Stevens}}, \bibinfo {author} {\bibfnamefont {Q.}~\bibnamefont
  {Zhou}}, \bibinfo {author} {\bibfnamefont {S.}~\bibnamefont {Grossmann}},
  \bibinfo {author} {\bibfnamefont {R.}~\bibnamefont {Verzicco}}, \bibinfo
  {author} {\bibfnamefont {K.-Q.}\ \bibnamefont {Xia}}, \ and\ \bibinfo
  {author} {\bibfnamefont {D.}~\bibnamefont {Lohse}},\ }\href@noop {}
  {\bibfield  {journal} {\bibinfo  {journal} {Phys. Rev. E},\ }\textbf
  {\bibinfo {volume} {85}},\ \bibinfo {pages} {027301} (\bibinfo {year}
  {2012})}\BibitemShut {NoStop}%
\bibitem [{\citenamefont {Brown}\ and\ \citenamefont
  {Ahlers}(2007)}]{BA07_EPL}%
  \BibitemOpen
  \bibfield  {author} {\bibinfo {author} {\bibfnamefont {E.}~\bibnamefont
  {Brown}}\ and\ \bibinfo {author} {\bibfnamefont {G.}~\bibnamefont {Ahlers}},\
  }\href@noop {} {\bibfield  {journal} {\bibinfo  {journal} {Europhys. Lett.},\
  }\textbf {\bibinfo {volume} {80}},\ \bibinfo {pages} {14001} (\bibinfo {year}
  {2007})}\BibitemShut {NoStop}%
\bibitem [{\citenamefont {Weiss}\ and\ \citenamefont {Ahlers}(2011)}]{WA11a}%
  \BibitemOpen
  \bibfield  {author} {\bibinfo {author} {\bibfnamefont {S.}~\bibnamefont
  {Weiss}}\ and\ \bibinfo {author} {\bibfnamefont {G.}~\bibnamefont {Ahlers}},\
  }\href@noop {} {\bibfield  {journal} {\bibinfo  {journal} {J. Fluid Mech.},\
  }\textbf {\bibinfo {volume} {676}},\ \bibinfo {pages} {5} (\bibinfo {year}
  {2011})}\BibitemShut {NoStop}%
\bibitem [{\citenamefont {Grossmann}\ and\ \citenamefont {Lohse}(2011)}]{GL11}%
  \BibitemOpen
  \bibfield  {author} {\bibinfo {author} {\bibfnamefont {S.}~\bibnamefont
  {Grossmann}}\ and\ \bibinfo {author} {\bibfnamefont {D.}~\bibnamefont
  {Lohse}},\ }\href@noop {} {\bibfield  {journal} {\bibinfo  {journal} {Phys.
  Fluids},\ }\textbf {\bibinfo {volume} {23}},\ \bibinfo {pages} {045108}
  (\bibinfo {year} {2011})}\BibitemShut {NoStop}%
\bibitem [{\citenamefont {{von K\'arm\'an}}(1930)}]{Ka30}%
  \BibitemOpen
  \bibfield  {author} {\bibinfo {author} {\bibfnamefont {T.}~\bibnamefont {{von
  K\'arm\'an}}},\ }\href@noop {} {\bibfield  {journal} {\bibinfo  {journal}
  {Nachr. Ges. Wiss. G\"ottingen, Math.-Phys. Kl.},\ }\textbf {\bibinfo
  {volume} {58-76}},\ \bibinfo {pages} {322} (\bibinfo {year}
  {1930})}\BibitemShut {NoStop}%
\bibitem [{\citenamefont {Prandtl}(1932)}]{Pr32}%
  \BibitemOpen
  \bibfield  {author} {\bibinfo {author} {\bibfnamefont {L.}~\bibnamefont
  {Prandtl}},\ }\href@noop {} {\bibfield  {journal} {\bibinfo  {journal}
  {Ergeb. Aerodyn. Versuch, G\"ottingen},\ }\textbf {\bibinfo {volume} {IV}},\
  \bibinfo {pages} {18} (\bibinfo {year} {1932})}\BibitemShut {NoStop}%
\bibitem [{\citenamefont {Marusic}\ \emph {et~al.}(2010)\citenamefont
  {Marusic}, \citenamefont {McKeon}, \citenamefont {Monkewitz}, \citenamefont
  {Nagib}, \citenamefont {Smits},\ and\ \citenamefont
  {Sreenivasan}}]{MMMNSS10}%
  \BibitemOpen
  \bibfield  {author} {\bibinfo {author} {\bibfnamefont {I.}~\bibnamefont
  {Marusic}}, \bibinfo {author} {\bibfnamefont {B.~J.}\ \bibnamefont {McKeon}},
  \bibinfo {author} {\bibfnamefont {P.~A.}\ \bibnamefont {Monkewitz}}, \bibinfo
  {author} {\bibfnamefont {H.~M.}\ \bibnamefont {Nagib}}, \bibinfo {author}
  {\bibfnamefont {A.~J.}\ \bibnamefont {Smits}}, \ and\ \bibinfo {author}
  {\bibfnamefont {K.~R.}\ \bibnamefont {Sreenivasan}},\ }\href@noop {}
  {\bibfield  {journal} {\bibinfo  {journal} {Phys. Fluids},\ }\textbf
  {\bibinfo {volume} {22}},\ \bibinfo {pages} {065103} (\bibinfo {year}
  {2010})}\BibitemShut {NoStop}%
\bibitem [{\citenamefont {Stevens}\ \emph {et~al.}(2011)\citenamefont
  {Stevens}, \citenamefont {Lohse},\ and\ \citenamefont {Verzicco}}]{SLV11}%
  \BibitemOpen
  \bibfield  {author} {\bibinfo {author} {\bibfnamefont {R.~J. A.~M.}\
  \bibnamefont {Stevens}}, \bibinfo {author} {\bibfnamefont {D.}~\bibnamefont
  {Lohse}}, \ and\ \bibinfo {author} {\bibfnamefont {R.}~\bibnamefont
  {Verzicco}},\ }\href@noop {} {\bibfield  {journal} {\bibinfo  {journal} {J.
  Fluid Mech.},\ }\textbf {\bibinfo {volume} {688}},\ \bibinfo {pages} {31}
  (\bibinfo {year} {2011})}\BibitemShut {NoStop}%
\bibitem [{\citenamefont {Ahlers}\ \emph
  {et~al.}(2009){\natexlab{b}}\citenamefont {Ahlers}, \citenamefont
  {Funfschilling},\ and\ \citenamefont {Bodenschatz}}]{AFB09}%
  \BibitemOpen
  \bibfield  {author} {\bibinfo {author} {\bibfnamefont {G.}~\bibnamefont
  {Ahlers}}, \bibinfo {author} {\bibfnamefont {D.}~\bibnamefont
  {Funfschilling}}, \ and\ \bibinfo {author} {\bibfnamefont {E.}~\bibnamefont
  {Bodenschatz}},\ }\href@noop {} {\bibfield  {journal} {\bibinfo  {journal}
  {New J. Phys.},\ }\textbf {\bibinfo {volume} {11}},\ \bibinfo {pages}
  {123001} (\bibinfo {year} {2009}{\natexlab{b}})}\BibitemShut {NoStop}%
\bibitem [{\citenamefont {Brown}\ \emph {et~al.}(2005)\citenamefont {Brown},
  \citenamefont {Nikolaenko},\ and\ \citenamefont {Ahlers}}]{BNA05}%
  \BibitemOpen
  \bibfield  {author} {\bibinfo {author} {\bibfnamefont {E.}~\bibnamefont
  {Brown}}, \bibinfo {author} {\bibfnamefont {A.}~\bibnamefont {Nikolaenko}}, \
  and\ \bibinfo {author} {\bibfnamefont {G.}~\bibnamefont {Ahlers}},\
  }\href@noop {} {\bibfield  {journal} {\bibinfo  {journal} {Phys. Rev.
  Lett.},\ }\textbf {\bibinfo {volume} {95}},\ \bibinfo {pages} {084503}
  (\bibinfo {year} {2005})}\BibitemShut {NoStop}%
\end{thebibliography}
%

\end{document}